\documentclass[sigconf]{acmart}

\usepackage{booktabs}
\usepackage{algorithm}
\usepackage{algpseudocode}
\usepackage{bbold}
\usepackage{enumitem}
\usepackage{subfigure}
\usepackage{colortbl}
\usepackage{setspace}
\usepackage{graphicx}

\newcommand\captionshrink{\vspace*{-0.75\baselineskip}}
\newcommand\shrink{\vspace*{-0.5\baselineskip}}
\newcommand\minishrink{\vspace*{-0.15\baselineskip}}

\usepackage{enumitem}
\setlist{leftmargin=5mm}

\newtheoremstyle{mydef}
{2ex}
{2ex}
{\itshape}
{}
{\scshape}
{: }
{0.5em}
{}
\theoremstyle{mydef}

\renewenvironment{quote}{%
   \list{}{%
     \leftmargin0.5cm  
     \rightmargin0cm
   }
   \item\relax
}
{\endlist}

\begin{document}

\fancyhead{}  

\copyrightyear{2020}
\acmYear{2020}
\setcopyright{acmcopyright}\acmConference[KDD '20]{Proceedings of the 26th ACM SIGKDD Conference on Knowledge Discovery and Data Mining}{August 23--27, 2020}{Virtual Event, CA, USA}
\acmBooktitle{Proceedings of the 26th ACM SIGKDD Conference on Knowledge Discovery and Data Mining (KDD '20), August 23--27, 2020, Virtual Event, CA, USA}
\acmPrice{15.00}
\acmDOI{10.1145/3394486.3403202}
\acmISBN{978-1-4503-7998-4/20/08}

\title{Evaluating Conversational Recommender Systems\\ via User Simulation}

\author{Shuo Zhang}\authornote{Work done while at the University of Stavanger, Norway.}
\affiliation{%
  \institution{Bloomberg}
  \city{London}
  \country{United Kingdom}
}
\email{szhang611@bloomberg.net}

\author{Krisztian Balog}
\affiliation{%
  \institution{University of Stavanger}
  \city{Stavanger}
  \country{Norway}
}
\email{krisztian.balog@uis.no}

\begin{abstract}
Conversational information access is an emerging research area.  
Currently, human evaluation is used for end-to-end system evaluation, which is both very time and resource intensive at scale, and thus becomes a bottleneck of progress. 
As an alternative, we propose automated evaluation by means of simulating users.
Our user simulator aims to generate responses that a real human would give by considering both individual preferences and the general flow of interaction with the system.
We evaluate our simulation approach on an item recommendation task by comparing three existing conversational recommender systems.  We show that preference modeling and task-specific interaction models both contribute to more realistic simulations, and can help achieve high correlation between automatic evaluation measures and manual human assessments.
\end{abstract}

\ccsdesc[500]{Information systems~Users and interactive retrieval}
\ccsdesc[500]{Information systems~Recommender systems}
\ccsdesc[300]{Human-centered computing~HCI design and evaluation methods}

\keywords{User simulation; conversational recommendation; conversational information access}

\maketitle

\section{Introduction}
\minishrink

Conversational information access is a newly emerging research area that aims at providing access to digitally stored information over a dialog interface~\citep{Trippas:2019:SCS,Vakulenko:2019:KCS}. 
It is specifically concerned with a goal-oriented sequence of exchanges, including complex information seeking and exploratory information gathering, and multi-step task completion and recommendation~\citep{Culpepper:2018:RFI}.
In this paper, we focus on the problem of item recommendation.  The conversational paradigm is particularly suited for this task, as it allows people to 
disclose their preferences~\citep{Christakopoulou:2016:TCR},
efficiently explore the search space~\citep{Zhang:2018:TCS},
and provide fine-grained feedback~\citep{Bi:2019:CPS}.
More specifically, we address the problem of evaluating  conversational recommender systems (referred to as \emph{conversational agents}).

Test-collection based evaluation has a large history in information retrieval (IR)~\citep{Sanderson:2010:TCB}, but it has limitations.  
It is possible to create an offline test collection for conversational agents to select the best response, in answer to a user utterance, from a set of possible candidates.  Assuming that the candidate generation step has been addressed by a different component, such reusable test collection would enable the comparison of different response ranking methods.  This is exactly the approach taken by the TREC Conversational Assistance Track benchmark initiative~\citep{TRECCAST} and also by others~~\citep{Aliannejadi:2019:ACQ}.  
However, this assessment is limited in scope to a single turn in a conversation; it does not tell us anything about the overall usefulness of the system or about users' satisfaction with the flow of the dialogue.
Collecting and annotating entire conversations is an option, but it is expensive, time-consuming, and does not scale.  Importantly, it would not yield a reusable test collection.
The evaluation of conversational information access systems, therefore, represents an open challenge and calls for additional methodologies to be considered.  Possible alternatives include laboratory user studies~\citep{Kelly:2009:MEI}, online evaluation~\citep{Hofmann:2016:OEI}, and simulated users~\citep{Maxwell:2019:MSS}.  Of these, we will be exploring user simulation in this work.

Our objective is to develop a user simulator that is (1) capable of producing responses that a real user would give in a certain dialog situation~\citep{Schatzmann:2006:SSU}, and (2) would enable to compute an automatic assessment of a conversational agent such that it is predictive of its performance with real users. 
We wish to accomplish this without making specific assumptions about the inner workings of conversational agents.  That is, we treat them much like black boxes.  
We further wish the simulator to be data driven, such that it can be used with any conversational agent only by supplying a small corpus of annotated dialogues real users have conducted with the agent. 

We build on the well-established Agenda-based User Simulator~\citep{Schatzmann:2007:AUS} as our general framework, and explore multiple options for modeling interactions and user preferences.  Specifically, we develop a task-specific interaction model to more directly capture the flow of the conversational item recommendation task.
For preference modeling, we present an approach for ensuring the consistency of responses based on the notion of personal knowledge graphs~\citep{Balog:2019:PKG}.
We evaluate our simulation approaches by comparing three existing conversational movie recommender systems, using both automatic and manual evaluation.  We find that more sophisticated interaction and preference modeling leads to more realistic simulation, and we achieve high overall correlation with real users.

In summary, this paper makes the following novel contributions:
\begin{itemize}	
	\item A general framework for evaluating conversational recommender agents via simulation.
	\item Interaction and preference models to better control the conversation flow and to ensure the consistency of responses given by the simulated user.
	\item An experimental comparison of three conversational movie recommender agents, using both real and simulated users.   
	\item An analysis of comments collected from human evaluation, and identification of areas for future development.
\end{itemize}
Our simulation platform is made publicly available.\footnote{\url{https://github.com/iai-group/kdd2020-usersim}}

\section{Related Work}
\label{sec:rw}
\minishrink

Our work investigates how to evaluate conversational recommender systems via user simulation, and is located in the intersection dialogue systems, conversational information access, and evaluation.  

\shrink
\subsection{Dialogue Systems}
\minishrink
Dialogue systems communicate with users in natural language (text, speech, or both).
They can be broadly categorized into two groups: \emph{non-task-oriented systems} (also known as \emph{chatbots}) and \emph{task-oriented systems}~\citep{Chen:2017:SDS,Serban:2018:SAC}.
Chatbots aim to carry on an extended conversation (``chit-chat'') with the goal of mimicking unstructured human-human interactions. 
Task-oriented systems, on the other hand, aim to assist users to complete some specific task (e.g., give navigation directions, control appliances, book a flight, buy a product, etc.).  Our work falls in this latter category.
Modern task-oriented dialogue systems are based on a \emph{dialogue-state} (or \emph{belief-state}) architecture~\citep{Jurafsky:2019:SLP}, capitalizing on the notion of \emph{dialogue acts} (i.e., task-specific intents that are being communicated).

There is a long history of utilizing user simulation in the context of spoken dialog systems~\citep{Schatzmann:2006:SSU}. 
Simulation is mainly used for dialogue policy learning and end-to-end dialogue training, in order to reduce time and effort by generating large-scale utterances of real users~\citep{Schatzmann:2006:SSU}. 
Early work can be categorized into rule-based~\citep{Chung:2004:DFS} and corpus-based methods~\citep{Griol:2013:ADS, Schatzmann:2007:AUS}.
Recent works employ neural approaches, esp. sequence-to-sequence models~\citep{Kreyssig:2018:NUS,Asri:2016:ASM}.
The most widely used approach for policy optimization is the Agenda-Based User Simulator~\citep{Schatzmann:2007:AUS}, which represents the user state as a stack of user actions, called the agenda.  Our work also builds on this method.
Simulation can also be used to evaluate different aspects of a dialogue system~\citep{Griol:2013:ADS}, which is our focus in this paper.

\shrink
\subsection{Conversational Information Access}
\minishrink

Conversational information access is concerned with a goal-oriented sequence of exchanges~\citep{Culpepper:2018:RFI}, where the agent aims to help the user to satisfy their information need, by supporting them in finding, exploring, and understanding the possible options and information objects that are available~\citep{Leif:2018:CAI}.
When resolving information needs, the conversational agent should consider both short- and long-term knowledge of the user~\citep{Radlinski:2017:TFC}.
Recently, progress has been made on specific subtasks for conversational information access, including response ranking~\citep{Yang:2018:RRD}, asking clarifying questions~\citep{Aliannejadi:2019:ACQ}, predicting user intent~\citep{Qu:2019:UIP}, and preference elicitation~\citep{Bi:2019:CPS,Christakopoulou:2016:TCR}.
End-to-end evaluation, however, has received little attention to date, due to the lack of appropriate evaluation resources and methodology. With this paper, we aim to start filling this gap.

\shrink
\subsection{Evaluation}
\minishrink

Conversational recommenders follow a task-oriented dialog system architecture, consisting of natural language understanding (NLU), natural language generation (NLG), and dialog manager (DM). 
Evaluation may be performed on the component-level or end-to-end.

Component-level evaluation has primarily focused on NLU and NLG.
NLU is often viewed as a classification task and is evaluated in terms of precision, recall, and F1-score~\citep{Papangelis:2019:CMD,Asri:2016:ASM} or intent/slot error rates~\citep{Li:2017:ETC}.
NLG is commonly assessed using word overlap-based metrics from machine translation, such as BLEU, METEOR, and ROUGE~\citep{Belz:2006:CAA,Papangelis:2019:CMD}. 
These metrics, however, turn out to correlate very poorly with human judgments, due to the many possible responses to any given turn~\citep{Liu:2016:HTE}.
An alternative is to consider the meaning of each word by using  embedding-based metrics~\citep{Liu:2016:HTE}.  
In addition to automatic means of evaluation, where all the above metrics fail, human evaluation is also considered.  
For example, \citet{Belz:2006:CAA} use NIST, BLEU, and ROUGE for automatic evaluation, and a 6-point scale for human evaluation. 
They find that the automatic metrics can be expected to correlate well with human judgments only if the reference texts used are of high quality.

End-to-end evaluation assesses the dialogue quality based on the generated dialogues. 
Metrics include but not limited to success rate, reward, and average dialogue turns~\citep{Papangelis:2019:CMD, Peng:2018:DIP}.  We also use these metrics in our evaluation.
Human evaluation has also been performed in terms of success rate~\citep{Peng:2018:DIP} and slot errors~\citep{Li:2017:ETC}.
A recently proposed alternative is \emph{adversarial evaluation}~\citep{Li:2017:ALN}.  Inspired by the Turing test, a classifier is trained to distinguish between human-generated and machine-generated responses; the more successful a system is at ``fooling'' the classifier, the better it is.  We perform a similar evaluation, but we ask crowd workers to try to perform this classification between real and simulated users.

\begin{table*}[t]
	\caption{Agent and user actions considered in this paper. Main actions are boldfaced.}
	\captionshrink
	\begin{tabular}{lll}
	\toprule
	\textbf{Category} & \textbf{Agent} & \textbf{User} \\
	\midrule
	\emph{Query Formulation}
		& \textbf{Reveal}: Disclose, Non-disclose, Revise, Refine, Expand
		& \textbf{Inquire}: Elicit, Clarify \\
	\midrule
	\emph{Set Retrieval}
		& \textbf{Inquire}: List, Compare,  Subset, Similar, 
		& \textbf{Reveal}: Show, List, Similar, Subset \\ 
		& \textbf{Navigate}: Repeat, Back, More, Note, Complete
		& \textbf{Traverse}: Repeat, Back, More, Record, End \\
	\midrule
	\emph{Mixed Initiative}
		& \textbf{Interrupt}, \textbf{Interrogate} 
		& \textbf{Suggest} \\	
	\bottomrule
	\end{tabular}
\label{tbl:actions}
\end{table*}

\minishrink
\section{Problem Statement}
\label{sec:ps}
\minishrink

Our goal is to develop an approach for evaluating conversational recommender systems (agents) via simulated users.
We specify two main requirements for the user simulator.
First, is should be capable of producing responses that a real user would give in a certain dialog situation~\citep{Schatzmann:2006:SSU}. 
Specifically, this entails (R1) generating responses that are \emph{consistent} with users' \emph{preferences}, and (R2) being able to \emph{follow} a task specific \emph{dialog flow}. 
Second, the simulator should (R3) enable to compute an \emph{automatic assessment} of the agent such that it is \emph{predictive} of its performance with real users. Formally: 
\begin{quote}
	For a given	system $S$ and user population $U$, the goal of user simulation $U^*$ is to predict the performance of $S$ when used by $U$, denoted as $M(S,U$).  For two systems $S_1$ and $S_2$, $U^*$ should be such that if $M(S_1,U) < M(S_2,U)$ then $M(S_1,U^*) < M(S_2,U^*)$.
\end{quote}

\shrink
\subsection{Simulation Framework}
\label{sec:ps:framework}
\minishrink

\begin{figure}[t]
   \centering
   \vspace*{-1\baselineskip}
   \includegraphics[width=0.45\textwidth]{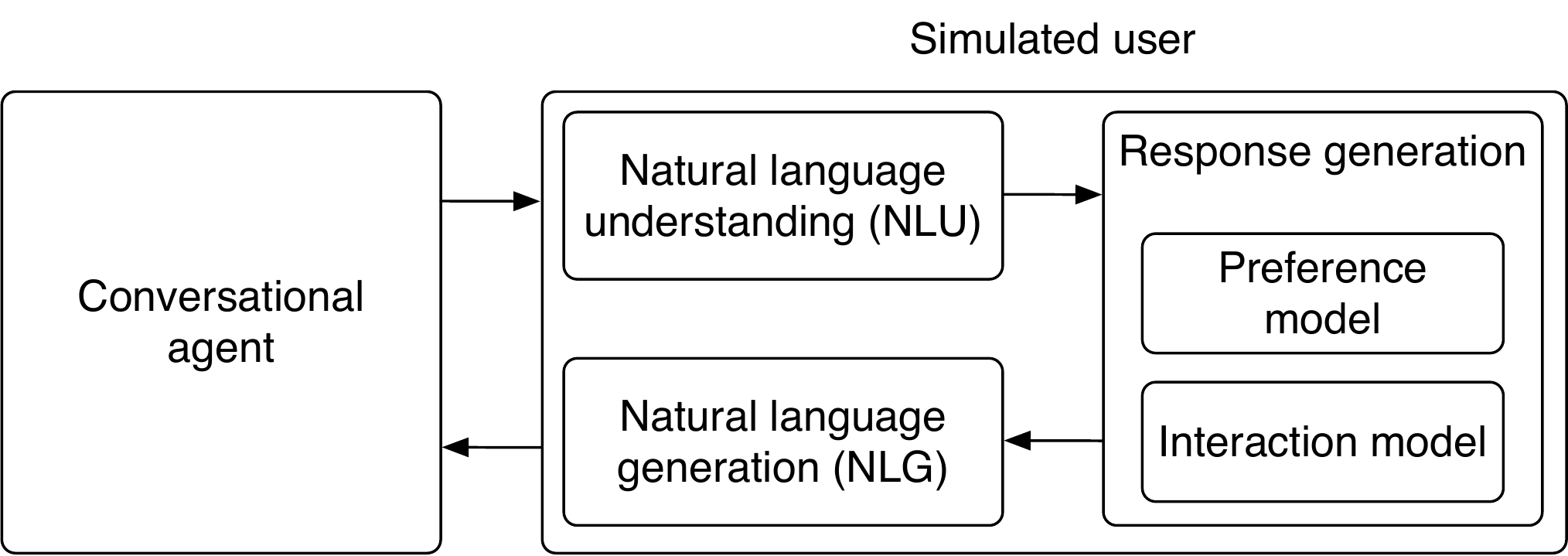} 
   \shrink
   \caption{Architecture of our user simulator.}
   \label{fig:arch}
   \vspace*{-1\baselineskip}
\end{figure}

The user simulator consists of the following main components, which are illustrated in Fig.~\ref{fig:arch}: natural language understanding, response generation, and natural language generation.  Our main focus in this paper is on the response generation part. 

\emph{Natural language understanding} is the task of translating an agent utterance into a structured format.
We will assume that the user has the capacity to perfectly understand the intent behind the agent's utterances. 
\emph{Response generation} is concerned with determining the next user action based on the understanding of the system's utterance.  
To ensure the consistency of user preferences (R1), we employ a \emph{preference model}, which is a structured representation of item- and set-level user preferences, based on the notion of a personal knowledge graph.
To follow a task-specific dialog flow (R2), we utilize an \emph{interaction model}.  We will assume that the user has some expectations regarding how the agent should act (i.e., we impose a pre-defined interaction policy).
\emph{Natural language generation} is the process of turning a structured response representation into natural language. 
We will use a simple template-based approach to turn structured intent representations into natural language text. 

\minishrink
\section{Modeling Simulated Users}
\label{sec:modeling}
\minishrink

Our objective is to simulate users for a specific task: conversational item recommendation.
Modeling dialogue as a Markov Decision Process, we employ agenda-based simulation~\citep{Schatzmann:2007:AUS} as the overall simulation framework (Sect.~\ref{sec:modeling:agenda}).
It operates on the notion of \emph{dialogue acts}, referred to as \emph{actions} henceforth, which represent task-specific intents that are being communicated in utterances.  Regarding the choice of actions, we take a subset of actions identified in~\citep{Leif:2018:CAI} and list them in Table~\ref{tbl:actions}.
To operationalize the agenda-based framework, we need a model of interaction that guides the simulated user through the conversation, i.e., helps to determine how to respond.
We consider both an existing general-purpose model and introduce a task-specific alternative (Sect.~\ref{sec:modeling:interaction}).
Furthermore, we need to model the preferences of the simulated user.  We present two alternatives, both of which hinge on the idea of sampling from historical user-item interactions (Sect.~\ref{sec:modeling:pref}).
Noting that this is not our focus, we detail natural language understanding and generation in Sect.~\ref{sec:modeling:nlug}.
Overall, our approach does not make any assumptions about the inner workings of the recommender agent.  We, however, assume that a small corpus of dialogs between humans and the agent, with turns annotated with user/agent actions, is available for training various components of the simulator.

\shrink
\subsection{Agenda-based Simulation}
\label{sec:modeling:agenda}
\minishrink

\begin{figure}[t]
   \centering
   \includegraphics[width=0.25\textwidth]{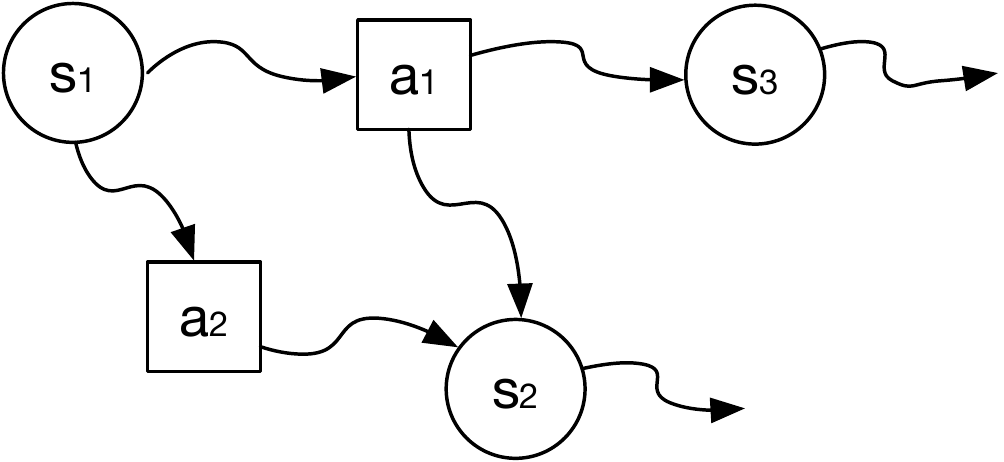} 
   \caption{Dialogue as a Markov Decision Process~\citep{Schatzmann:2006:SSU}.}
   \label{fig:mdp}
   \shrink
\end{figure}

Dialogue can effectively be modeled as a Markov Decision Process (MDP)~\citep{Schatzmann:2006:SSU}.
Every MDP is formally described by a finite space $\mathcal{S}$, a finite action set $\mathcal{A}$, and a set of transition probabilities.  At each time step (dialogue turn) $t$, the dialogue manager is in a particular state $s_t \in \mathcal{S}$.  By executing action $a_t \in \mathcal{A}$, it transitions into the next state $s_{t+1}$ according to the transition probability $P(s_{t+1}|s_t,a_t)$.  The Markov property ensures that the state at time $t+1$ depends only on the state and action at time $t$:
\begin{equation*}
	P(s_{t+1}|s_t,a_t,s_{t-1},a_{t-1},\dots,s_0,a_0)=P(s_{t+1}|s_t,a_t) ~.
\end{equation*}
Using the MDP model of dialogue, the dialogue manager can be visualized as an agent traveling through a network of dialogue states; see Fig.~\ref{fig:mdp}.

\begin{figure*}[ht]
   \centering
   \includegraphics[width=0.9\textwidth]{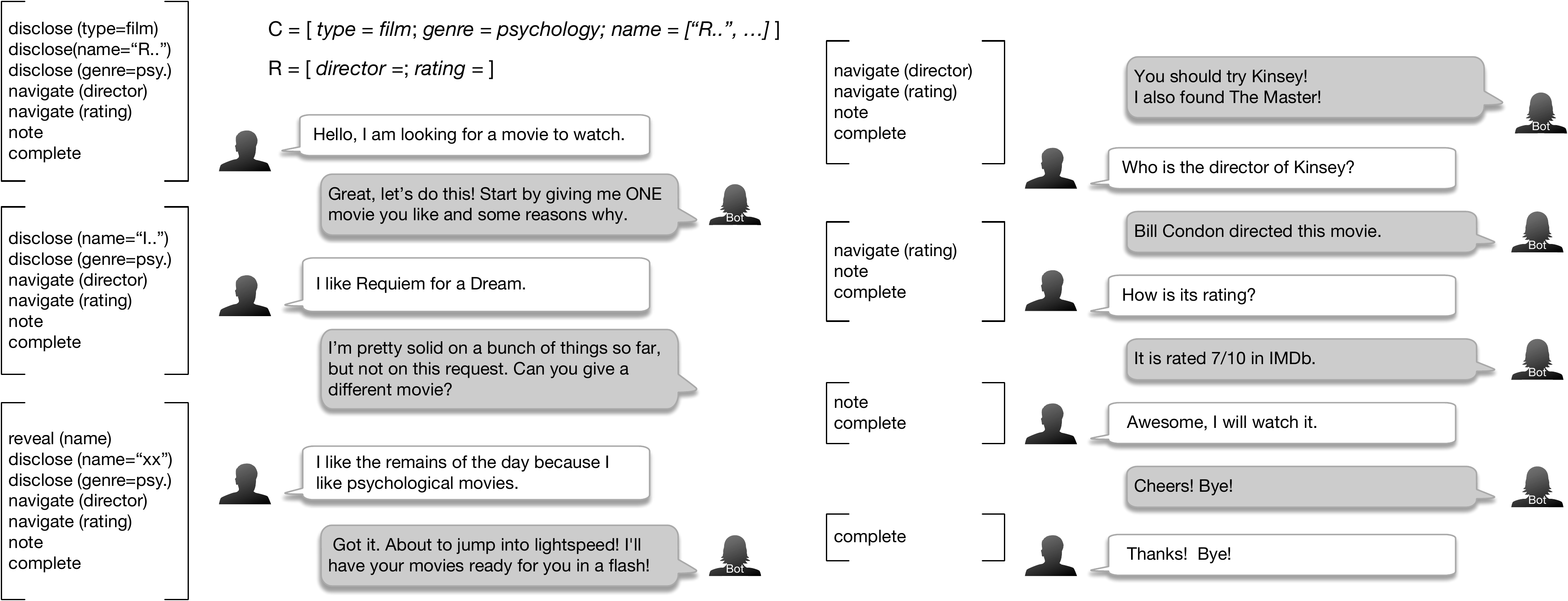} 
   \shrink
   \caption{Example dialogue with agenda sequence and state transition. The agenda is shown in square brackets.  The third agenda is a result of a push operations, all other agendas updates are pull operations.}
   \label{fig:dia}
\end{figure*}

The agenda-based simulator~\citep{Schatzmann:2007:AUS} provides a probabilistic method for bootstrapping the MDP dialogue process. 
The user state $s$ is factorized into action agenda $A$ and information-seeking goal $g$. $A$ is a stack-like representation for user actions that is dynamically updated; the next user action is selected from the top of the agenda. 
Specifically, the user agenda $A$ is a stack-like structure of length $n$, containing the user dialogue actions, where $A[1]$ denotes the bottom and $A[n]$ denotes the top item.
The update of $A$ is coherent to the state; new actions are pushed onto the agenda, and no longer relevant ones are removed.
We use $s_t$ and $s_{t+1}$ to represent two consecutive states in the diagram, and $a_t$ to represent the user action selected from $A_t$. $A_{t+1}$ is the agenda after taking $a_t$, and it is derived from $A_t$ for simplification. 
This way we have formalized the conversation into a sequence of states;  Fig.~\ref{fig:dia} illustrates state transitions via an example.
Accordingly, the estimation of the probability of going from one state to the next goes as follows: 
\begin{equation*}
	P(s_{t+1}|s_t, A_t) = P(A_{t+1}|A_t, g_{t+1}) \cdot P(g_{t+1}|A_t, g_t),
\end{equation*}
where $P(A_{t+1}|A_t, g_{t+1})$ is the agenda update and $P(g_{t+1}|A_t, g_t)$ is the goal update.
Goal $g$ is further decomposed into constrains $C$, specifying the type of information sought, and requests $R$, which specify the additional pieces of information requested from the agent.  We construct $g$ based on the preference model (to be detailed in Sect.~\ref{sec:modeling:interaction}).
The goal update is formalized as follows:
\begin{equation*}
	P(g_{t+1}|A_t, g_t) = P(R_{t+1}|A_t, R_t, C_{t+1}) \cdot P(C_{t+1}|A_t, R_t, C_t) \cdot \delta(g_{t+1}, g_t).
\end{equation*}
As suggested by~\citet{Schatzmann:2007:AUS}, the goal update may be simplified by hand-crafted heuristics.  Our heuristic is to check whether the agent understands the user action and gives the corresponding response. 
Thus, we base it only on an indicator function $\delta$:
\begin{equation*}
	P(g_{t+1}|A_t, g_t) = \delta(g_{t+1}, g_t) ~,
\end{equation*}
where $\delta(g_{t+1}, g_t)$ returns $1$ if the goal $g_t$ was accomplished and otherwise returns 0. 

Agenda updates are regarded as a sequence of pull or push operations, where dialogue actions are removed from or added to the top. 
An accomplished goal ($\delta(g_{t+1},g_t)\!=\!1$) indicates a pull, otherwise push.
For pull, the state transition probability simplifies to:
\begin{equation}
	P(s_{t+1}|A_t, s_t) = P(A_{t+1}|A_t, g_{t+1}) ~.\label{eq:pull}
\end{equation}
For the push operation, we need to find a replacement action $\widetilde{a}_{t}$, which remains to have the same goal as the original action $a_t$.  
The state transition probabilities are then computed according to:
\begin{equation}
	P(A_{t+1}|A_t, g_{t+1}) 
	= P(\widetilde{a}_{t}|A_t, g_{t+1}) ~. \label{eq:push} 
\end{equation}
The agenda updates, namely, the pull operation ($P(A_{t+1}|A_t, g_{t+1})$) and finding the replacement action in case of a push operation ($P(\widetilde{a}_{t}|A_t, g_{t+1})$) are informed by the interaction model, and will be detailed in the next subsection.

To sum up, we switch between pull and push (replace) operations by checking if the user action is met with an appropriate agent response. 
The dialogue is terminated when the agenda is empty.

\shrink
\subsection{Interaction Model}
\label{sec:modeling:interaction}
\minishrink

The interaction model defines how the agenda should be initialized ($A_0$) and updated ($A_t \rightarrow A_{t+1}$) throughout the conversation.  
We consider two interaction models: (1) an existing general-purpose conversational interaction model, QRFA, which will serve as our baseline, and (2) our model, CIR6, which is developed specifically for the conversational item recommendation task.
Before we detail these models, we need to specify the space of possible user actions.

\subsubsection{Action Space}

We base our user actions $\mathcal{A}$ on agent-human interactions for conversational search by \citet{Leif:2018:CAI}, which are listed below (with examples taken from~\citep{Leif:2018:CAI}). 
\begin{itemize}
	\item \textbf{Disclose}: The user expresses the information need either actively, or in response to the agent's question (``\emph{I would to arrange a holiday in Italy.}'').
	\item \textbf{Reveal}: It refers to the user revising, refining, or expanding constraints and requirements (``\emph{Actually, we need to go on the 3rd of May in the evening.}'' or ``\emph{Can you also check to see what kind of holidays are there available in Spain?}'').
	\item \textbf{Inquire}: Once the agents starts to show recommendations, the user may ask for related items (``\emph{Tell me about all the different things you can do in this place.}'') or ask for similar options (``\emph{What other regions in Europe are like that?}'').
	\item \textbf{Navigate}: In our definition, navigation entails both actions around navigating a list of recommendations (``\emph{Which one is the cheapest option?}'') as well as questions about a certain recommended item on the list (``\emph{What's the price of that hotel?}'').	
	\item \textbf{Note}: During the conversation, the user could mark or save specific items (``\emph{That hotel could be a possibility.}'' or ``\emph{Save that hotel for later.}'').
	\item \textbf{Complete}: Finally, the user can mark the end of the conversation (``\emph{Thanks for the help, bye.}'').
\end{itemize}
Note that we only use user actions to compose the agenda.  
That is, we generate the next action in the agenda directly based on the current user action, while treating the agent much like a black box.  We assume, however, that the simulator can ``understand'' a set of agent actions.  Specifically, we consider the agent actions listed in Table~\ref{tbl:actions} (for a detailed description of each, we refer the reader to~\citep{Leif:2018:CAI}). 
The NLU is trained to recognize this set of agent actions.  Then, at each turn, the agenda-based simulator can determine whether the agent responds to the user with an appropriate action (as captured by the indicator function $\delta$).  For example, an \emph{Inquire} user action can accept \emph{List} or \emph{Elicit} as an agent response; the full mapping is excluded due to space constraints and will be made available online. 

\subsubsection{QRFA Model}
\label{sec:modeling:qrfa}

\begin{figure}[t]
   \centering
   \includegraphics[width=0.35\textwidth]{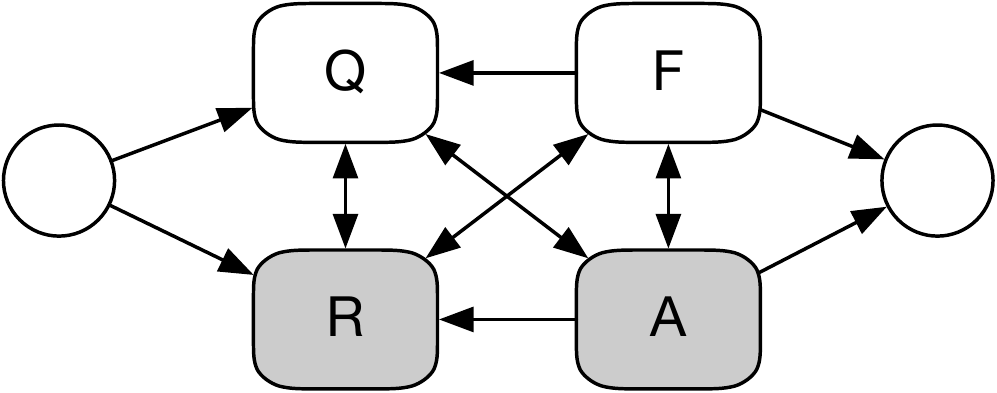} 
   \minishrink
   \caption{The QRFA model~\citep{Vakulenko:2019:QAD}. Agent actions are in grey.}
   \label{fig:qrfa}
   \shrink
\end{figure}

QRFA (Query, Request, Feedback, and Accept)~\citep{Vakulenko:2019:QAD} is a general model for conversational information seeking processes. 
It uses a simple schema for annotating utterances, with four basic classes: two for user (\emph{Query} and \emph{Feedback}) and two for agent (\emph{Request} and \emph{Answer}); see Fig.~\ref{fig:qrfa}.
\citet{Vakulenko:2019:QAD} use this model to discover frequent sequence patterns in dialogs with the help of process mining techniques.
QRFA provides good flexibility and generalizability to a wide number of use cases.
However, we need to make some adjustments before it can be applied in our scenario. 
First, for simulation purposes, where we are only interested in the user side, which has only two high-level classes (\emph{Query} and \emph{Feedback}).  We subdivide the high-level QRFA categories into our more fine-grained set of actions, as shown in Table~\ref{tbl:actions_qrfa}. 
Second, agenda initialization is a reverse process to pattern discovery, and there is a lack of methods. Therefore, we take an initial agenda $A_0$ by sampling from an annotated training corpus of human-agent conversations.  When estimating the state transition probabilities, we leverage the agent action $b_t$ (which is either \emph{Request} or \emph{Action}) that happens between two consecutive user actions $a_t$ and $a_{t+1}$ as a two-step transition probability.  
The transition probability matrix is estimated based on the training corpus. The transition probability between actions is then defined as follows:
\begin{equation*}
	P(A_{t+1}|A_t, g_{t+1}) = 
	\begin{cases}
		P(A_{t+1}|b_t)P(b_t|A_t)& \delta(g_{t+1}, g_t)=1\\
		P(\widetilde{a}_{t}|b_t) & \delta(g_{t+1}, g_t)=0,
	\end{cases}
	\label{eq:sr}
\end{equation*}
where $\delta(g_{t+1}, g_t)$ indicates whether $b_t$ responds to $a_t$ with an appropriate action.  If yes, then we perform a pull operation and remove $a_t$ from the agenda. 
Otherwise, it is a push operation, where a replacement action $\widetilde{a}_t$ is sampled based on the last agent action $b_t$.
Mind that the agenda updates are performed on the course-grained level (i.e., only Query and Feedback actions).  We then probabilistically sample a corresponding fine-grained (sub)action (cf. Table~\ref{tbl:actions_qrfa}) based on historical dialogs.

\begin{table}[t]
	\caption{Mapping the action set used in this paper to high-level QRFA categories.}
	\captionshrink
	\small
	\begin{tabular}{@{~}lp{6.7cm}}
	\toprule
	\textbf{Category} & \textbf{Actions} \\
	\midrule
	Query & Reveal, Disclose, Non-disclose, Revise, Refine, Expand, Inquire, List, Compare, Subset, Similar, Navigate, Repeat, Interrupt, Interrogate \\
	Request & Inquire, Elicit, Clarify, Suggest \\
	Feedback & Back, More, Note, Complete \\
	Answer & Show, List, Similar, Subset, Repeat, Back, More, Record, End \\
	\bottomrule
	\end{tabular}
\label{tbl:actions_qrfa}
\end{table}

\subsubsection{CIR6 Model}
\label{sec:modeling:cir6}

\begin{figure}[t]
   \centering
   \includegraphics[width=0.4\textwidth]{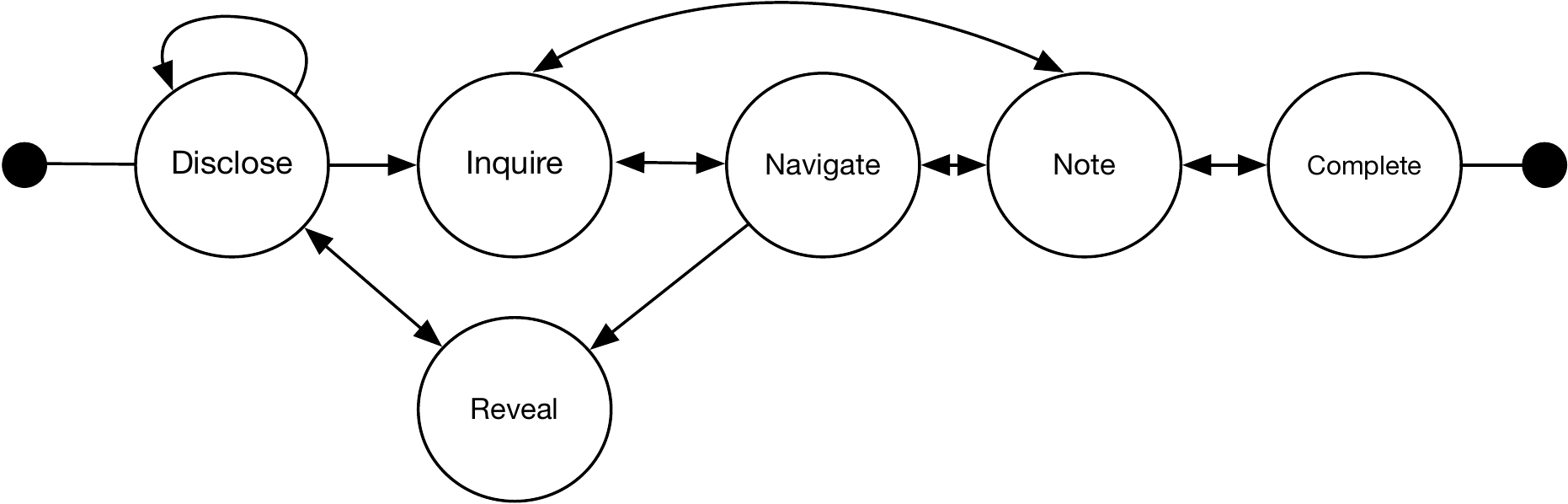} 
   \minishrink
   \caption{State diagram of our CIR6 model. Different from QRFA, we model only user actions as states.}
   \label{fig:sda}
   \shrink
   \shrink
\end{figure}

Next, we present our interaction model, which is designed to more directly capture the flow of the conversational item recommendation task.  It considers six main user action, hence it is termed CIR6 (for Conversational Item Recommendation).  Figure~\ref{fig:sda} presents the state diagram.
Using this model, we can generate the next action in the agenda directly based on the current user action, without having to resort to transition probability estimations for agent actions.  Formally:
\begin{equation*}
	P(A_{t+1}|A_t, g_{t+1}) = 
	\begin{cases}
		P(A_{t+1}|A_t)\cdot \mathbb{1}(a_{t+1}, a_t)& \delta(g_{t+1}, g_t)=1\\
		P(\widetilde{a}_{t}|b_t) & \delta(g_{t+1}, g_t)=0,
	\end{cases}
	\label{eq:sr}
\end{equation*}
where $\mathbb{1}(a_{t+1}, a_t)$ indicates if the two consecutive actions are connected in the state diagram (Fig.~\ref{fig:sda}) or not.
We compute the conditional probability $P(A_{t+1}|A_t)$ based on action distributions in a training corpus, i.e., the number of times $a_{t}$ was followed by $a_{t+1}$.  


\shrink
\subsection{Preference Model}
\label{sec:modeling:pref}
\minishrink

The preference model is meant to capture individual differences and personal tastes.  Here, preferences are represented as a set of attribute-value pairs. 
We assume that a sufficiently large corpus of historical user-item interactions is available.  In order to create a realistic preference model, we first randomly choose a user from the corpus, then subsample from historical interactions of that user.  The set of sampled items is denoted as $I_u$.  We will assume that the simulated user has seen/consumed all items in this set.

\subsubsection{Single Item Preference}
\label{sec:modeling:single}

Recommender systems mostly elicit preferences by asking the user to provide one or more favored items (e.g., favorite movies).
Whenever a user is prompted whether they had seen/consumed a specific item $i$, we check if $i \in I_u$ an answer accordingly.  
When the user is prompted for their preference of a given item (e.g., ``\emph{Did you like it?}") we provide a positive/negative response by flipping a coin. 
This approach, therefore, offers limited consistency.  Items that are seen/consumed are rooted in real user behavior, but the preferences expressed about them are not.

\subsubsection{Personal Knowledge Graph}
\label{sec:modeling:pkg}

In order to have a more realistic model of user preferences, we build a personal knowledge graph (PKG)~\citep{Balog:2019:PKG}.
The PKG has two types of nodes: items and attributes.
For this approach, we will assume that the corpus of historical user-item interactions contains not only seen/consumed information, but also preferences (i.e., ratings).
We divide the $I_u$ into sets of liked and disliked items, $I_u^+$ and $I_u^-$, respectively, based on the ratings.
Given an attribute $j \in J$, we infer the rating for that attribute by considering the ratings of items that have that attribute:
\begin{equation*}
	r_j = \frac{1}{|I_j|}\sum_{i \in I_j} r_{i},
\end{equation*}
where $I_j$ denotes the set of items that have attribute $j$, and $r_i$ is the rating of the item $i$ ($r_i \in [-1,1]$).
We can then classify attributes into liked and disliked sets, $J_u^+$ and $J_u^-$, respectively.
Whenever the user is asked about preferences of a specific item or attribute, those answers are based on the PKG.  This ensures that all preference statements expressed by the simulated user will be consistent.

\shrink
\subsection{NL Understanding and Generation}
\label{sec:modeling:nlug}
\minishrink

\emph{Natural language understanding} (NLU) is responsible for annotating agent utterances with actions (according to Table~\ref{tbl:actions}) and entities (by linking them to a domain specific knowledge base).
To carry out these tasks, we assume that a small corpus of dialogs with the conversational agent is available, which is labeled with agent actions.  The size of the annotated corpus depends on the variety of language the agent uses.  For example, the agents we will consider in our evaluation use rather rigid patterns, therefore, only a limited amount of labeled data is required.  
We use a simple retrieval-based approach for NLU, where we identify the most similar utterance from the corpus for each input utterance, and take the corresponding action~\citep{Jurafsky:2019:SLP}.  For entity linking, we first extract patterns (templates) from the labeled corpus that contain placeholders for entity mentions.  Then, we use a retrieval-based approach for disambiguation based on surface forms alone.

\emph{Natural language generation} (NLG) is concerned with creating a textual user utterance based on the user action and associated slots.  We follow a template-based approach where, for each action, we randomly select from a small number of hand-crafted response variations.  To make it more human-like, a few of the templates purposefully contain typos.

\minishrink
\section{Evaluation Architecture}
\label{sec:em}
\minishrink

\begin{figure}[t]
   \centering
   \includegraphics[width=0.4\textwidth]{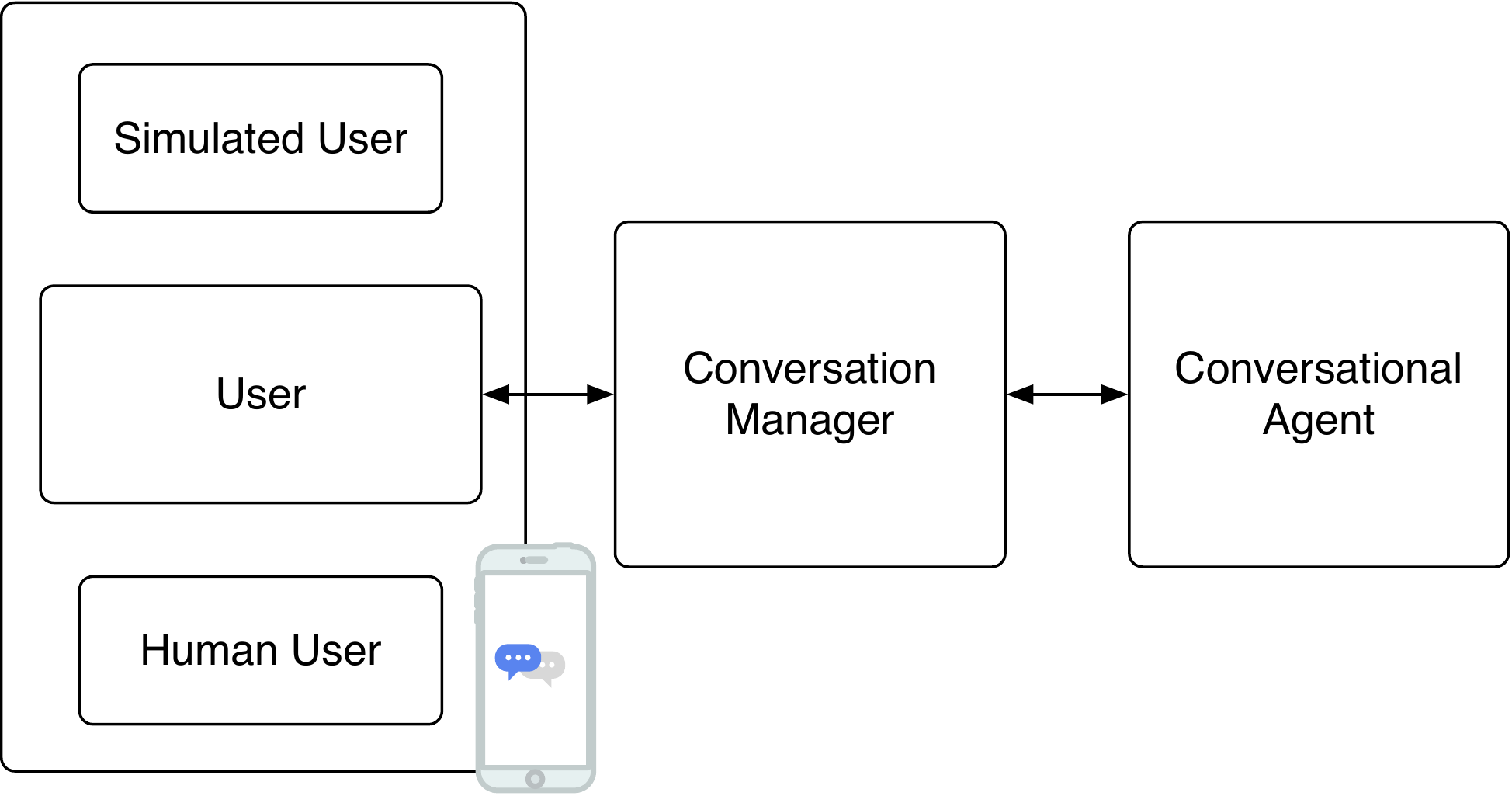} 
   \shrink
   \caption{Architecture of our evaluation platform.}
   \label{fig:frm}
   \shrink
\end{figure}

We evaluate conversational agents with both real and  simulated users.  This is facilitated by a conversation manager, which is a glue module connecting the agent and simulated/real users; see Fig.~\ref{fig:frm}.

\shrink
\subsection{Conversational Agents}
\label{sec:em:ca}
\minishrink

We consider three conversational agents for movie recommendation; two of these are existing third-party systems, while the third one was developed by us. 

\begin{itemize}
	\item \emph{And chill}\footnote{\url{http://www.andchill.io/}} is a single-purpose, consumer-oriented chatbot that a user can send messages to on Facebook and ask for a Netflix recommendation.  After answering a few questions such as a liked movie and the reason why liking it, the agent sends movie recommendations based on the user's preferences.
	\item \emph{Kelly Movie Bot}\footnote{\url{https://github.com/Sundar0989/Movie\_Bot}} is a simple bot that answers questions about a specific movie, such as rating, genre, and can also recommend similar movies.   The underlying data collection is the Kaggle Movies Recommender System dataset,\footnote{\url{https://www.kaggle.com/rounakbanik/movie-recommender-systems/data}} which is based on the MovieLens dataset.  The natural language components utilize the IBM Watson API services.\footnote{\url{https://www.ibm.com/cloud/watson-assistant/}}  We extended the original Kelly Movie Bot with a number of additional intents (to allow users to indicate their preferences and whether they have already watched a given movie).  
 
	\item \emph{Our movie recommender system} is based on the Plato Research Dialogue System.\footnote{\url{https://github.com/uber-research/plato-research-dialogue-system}} This agent can answer questions about movies (such as directors, summary, ratings, etc.) and can provide recommendations based on genres.  It also solicits feedback on movies the user has watched.
\end{itemize}
\noindent
For our experiments, we anonymized the three agents, by assigning labels \textbf{A}, \textbf{B}, and \textbf{C} to them in random order. 
All three agents support query formulation, set retrieval, and mixed-initiative properties.
For two of the agents we report results by averaging by 100 conversations,  
while for one of the agents we report only 25 conversations due to access restrictions.  

\shrink
\subsection{Simulated Users}
\label{sec:eval:data}
\minishrink

To instantiate a simulated user, we discuss how to initialize the preference model and train the interaction model; cf. Fig.~\ref{fig:arch}.  

To initialize the \emph{preference model} we utilize the MovieLens data\-set~\citep{Harper:2015:MDH}. 
A user $u$ in MovieLens reviews and rates movies from 0.5 to 5; we write $M_u$ to denote the set of movies reviewed. 
Items rated at least 4 are regarded as liked, items rated 2 or below are regarded as disliked, and the remaining ratings in between are treated as neutral.
For any simulated user, we construct the preference model by randomly sampling historical preferences of a real user from this dataset. Specifically, we sample 8 rated movies as items (of which at least one must be a liked item), and infer a personal knowledge graph (i.e., movie and genre preferences) from these items. 

The \emph{interaction model} is trained based on behaviors of real human users.  Specifically, we collect conversations between the three conversational agents and human users using a crowdsourcing platform (Amazon Mechanical Turk).  All conversational agents were deployed as a Telegram application.\footnote{\url{https://telegram.org/}} 
Crowd workers were instructed to find the respective channel, engage in a conversation with the agent, and keep interacting with the agent until they receive a movie recommendation they like.
For each conversational agent, we collected 25 successful dialogs (each with a different user) and paid \$1.5 for each conversation.  We regard a conversation successful if it covers the properties of both query formulation and set retrieval (mixed initiative is optional).
Then, the utterances in each conversation are annotated manually with the actions listed in Table~\ref{tbl:actions}.  For example, the system utterance ``Could you give me one movie you like?'' is labeled as \emph{Elicit}.
The annotations were performed by the paper's authors and disagreements were resolved through discussion.
The conditional probabilities in Sect.~\ref{sec:modeling:interaction} are estimated based on these empirical distributions.

\begin{table}[t]
	\caption{Simulation approaches used in our experiments. All use the same state transition modeling, NLU, and NLG.}
	\captionshrink
	\begin{tabular}{lll}
	\toprule
	 \textbf{Method} & \textbf{Interaction Model} & \textbf{Preference Model} \\
	\midrule
    QRFA-Single & QRFA (\S\ref{sec:modeling:qrfa}) & Single item  (\S\ref{sec:modeling:single}) \\
    CIR6-Single & CIR6 (\S\ref{sec:modeling:cir6}) & Single item  (\S\ref{sec:modeling:single}) \\
    CIR6-PKG & CIR6 (\S\ref{sec:modeling:cir6}) & PKG (\S\ref{sec:modeling:pkg}) \\
	\bottomrule
	\end{tabular}
	\shrink
\label{tbl:method}
\end{table}

\begin{table*}[t]
	\caption{Comparison of the characteristics of dialogs with real and simulated users, for different conversational agents (A--C).}
	\captionshrink	
	\begin{tabular}{l|rrr|rrr|rrr}
	\toprule
	& 
		\multicolumn{3}{c|}{\textbf{AvgTurns}} &
		\multicolumn{3}{c|}{\textbf{UserActRatio}} &
		\multicolumn{3}{c}{\textbf{DS-KL}} \\
	\textbf{Method} & 
		\multicolumn{1}{c}{\textbf{A}} & 
		\multicolumn{1}{c}{\textbf{B}} & 
		\multicolumn{1}{c|}{\textbf{C}} &
		\multicolumn{1}{c}{\textbf{A}} & 
		\multicolumn{1}{c}{\textbf{B}} & 
		\multicolumn{1}{c|}{\textbf{C}} &
		\multicolumn{1}{c}{\textbf{A}} & 
		\multicolumn{1}{c}{\textbf{B}} & 
		\multicolumn{1}{c}{\textbf{C}} \\
	\midrule
	Real users &  
		9.20 & 14.84  & 20.24 &
		0.374 & 0.501 & 0.500 &
		- & - & - \\
	QRFA-Single & 
		10.52 & 12.28 & 17.51 & 
		0.359 & 0.500 & 0.500 &
		0.027 & 0.056 & 0.029 \\
	CIR6-Single & 
		9.44 & 12.75 & 15.92 &
		0.382 & 0.500 & 0.500 &
		0.055 & 0.040 & 0.025 \\
	CIR6-PKG & 
		6.16 & 9.87 & 10.56 &
		0.371 & 0.500 & 0.500 &
		0.075 & 0.056 & 0.095 \\
	\bottomrule
	\end{tabular}
	\shrink
\label{tbl:characteristics}
\end{table*}

\begin{table*}[t]
	\caption{Performance of conversational agents using real vs. simulated users, in terms of Reward and Success Rate.  We show the relative ordering of agents (A--C), with evaluation scores in parentheses.}
	\captionshrink
	\begin{tabular}{l|c|c}
	\toprule
	\textbf{Method} & \textbf{Reward} & \textbf{Success Rate} \\  
	\midrule
	Real users &
		\textbf{A} (8.88) $>$ \textbf{B} (7.56) $>$ \textbf{C} (6.04) &
		\textbf{B} (0.864) $>$ \textbf{A} (0.833) $>$ \textbf{C} (0.727) \\
	QRFA-Single & 
		\textbf{A} (8.04) $>$ \textbf{B} (7.41) $>$ \textbf{C} (6.30) &
		\textbf{B} (0.836) $>$ \textbf{A} (0.774) $>$ \textbf{C} (0.718) \\
	CIR6-Single & 
		\textbf{A} (8.64) $>$ \textbf{B} (8.28) $>$ \textbf{C} (6.01) &
		\textbf{B} (0.822) $>$ \textbf{A} (0.807) $>$ \textbf{C} (0.712) \\
	CIR6-PKG & 
		\textbf{A} (11.12) $>$ \textbf{B} (10.65) $>$ \textbf{C} (9.31) &
		\textbf{A} (0.870) $>$ \textbf{B} (0.847) $>$ \textbf{C} (0.784) \\
	\bottomrule
	\end{tabular}
	\shrink
\label{tbl:auto}
\end{table*}

\minishrink
\section{Experimental Evaluation}
\label{sec:eval}
\minishrink

The main research question we seek to answer with our experiments is the following: \emph{Can simulation be used to predict the performance of a conversational recommender agent with real users?}
Each of the following subsections addresses a more specific sub-question.  

\noindent
Table~\ref{tbl:method} summarizes the three simulation approaches, which will be compared against real users, using both automatic and manual evaluation methods. 
Note that CIR6-PKG forces to empty the agenda once the user finds an item that they would like.  This can significantly reduce the number of turns taken with the agent.

\minishrink
\subsection{Characteristics of Conversations}
\minishrink

\emph{(RQ1) How well do our simulation techniques capture the characteristics of conversations?}
To answer this question, we consider three statistical measures from the literature:
(1) \textbf{AvgTurns}: the average number of dialogue turns~\citep{Liu:2016:HTE};
(2) \textbf{UserActRatio}: the ratio of user and system acts~\citep{Schatzmann:2005:QEU}, which is a measure of user participation;
(3) \textbf{DS-KL}: a dissimilarity metric based on Kullback-Leibler divergence, which can be regarded as a measure of dialogue style~\citep{Pietquin:2013:SME}:
		\begin{equation*}
			DS(P||Q) = \frac{D_{KL}(P||Q) + D_{KL}(Q||P)}{2} ~,
		\end{equation*}
where $P$ and $Q$ are probability distributions over the different user actions observed in simulated and real dialogs, respectively, and $D_{KL}$ is the KL-divergence between two distributions. Note that this is an unbounded metric; the closer the number to zero, the more similar two distributions are.

Table~\ref{tbl:characteristics} presents the results.
We find that the first two simulation approaches, QRFA-Single and CIR6-Single, resemble quite closely the characteristics of conversations with real users, in terms of all three metrics.
As expected, CIR6-PKG tends to have significantly shorter average conversation length, since it terminates the dialog as soon as the user finds a recommendation they like.  Because of this, the distribution of the actions is also reshaped, as witnessed by higher DS-KL scores.  Interestingly, this method is the closest to real humans in terms of user participation (UserActRatio).

\begin{table*}[t]
	\caption{Side-by-side comparison results, with human evaluators guessing which of two dialogs with a given conversational agent (A--C) was performed by a simulated user (Win) vs. a real one (Loss); a Tie is given when the evaluator could not decide.}
	\captionshrink	
	\begin{tabular}{llllllllllllllll}
	\toprule
	  & & \textbf{A} &&&& \textbf{B} &&&& \textbf{C} &&&& \textbf{All} \\  
	  \cline{2-4} \cline{6-8} \cline{10-12} \cline{14-16}
	 	  & \textbf{Win} & \textbf{Lose} & \textbf{Tie} &&\textbf{Win} & \textbf{Lose} & \textbf{Tie} && \textbf{Win} & \textbf{Lose} & \textbf{Tie}  && \textbf{Win} & \textbf{Lose} & \textbf{Tie}\\  
	\midrule
	 QRFA-Single & 20 & 39 & 16 && 22 & 33 & 20 && 19 & 43 & 13 && 61 (27\%) & 115 (51\%) & 49 (22\%) \\	
	 CIR6-Single & 27 & 30 & 18 && 23 & 33 & 19 && 26 & 40 & 9 && 76 (33\%) & 103 (46\%) & 46 (21\%) \\
	 CIR6-PKG  & 22 & 39 & 14 && 27 & 29 & 19 && 32 & 25 & 18 && 81 (36\%) & 93 (41\%) & 51 (23\%) \\	
	\bottomrule
	\end{tabular}
	\shrink
\label{tbl:human}
\end{table*}

\shrink
\subsection{Performance Prediction}
\minishrink

\emph{(RQ2) How well do the relative ordering of systems according to some measure correlate when using real vs. simulated users?}
To answer this question, we perform end-to-end evaluation using two automatic evaluation metrics that have been used in the literature to evaluate the performance of task-oriented conversational agents. 

\textbf{Reward}: Motivated by ABUS~\citep{Schatzmann:2007:AUS}, the reward function assigns \emph{Full} points (20) for successful task completion and \emph{Cost} (1) for every user turn. 
We equally assign points to functions of query formulations \emph{Disclose} and \emph{Refinement}, set retrieval actions \emph{Inquire} and \emph{Navigation}, and mixed-initiative (4 points for each). 
For agents not supporting any of these functions, we deduct the corresponding points from \emph{Full}. For example, if an agent does not support navigate in set retrieval, \emph{Full} is set to 16 points.
We deem two consecutive \emph{Repeat} actions as one turn given that some bots do not support multi-turn \emph{Navigation}.
In the end, the reward function is:
$\mathrm{Reward} = \mathrm{max}\{0, \emph{Full} - \emph{Cost} \cdot T\}$, 
where $T$ is the number of user turns. 

\textbf{Success Rate}: We measure success rate on the turn level~\citep{Peng:2018:DIP} based on the appropriateness of agent actions. I.e., if the agent returns a wrong action, we deem this turn as a failure.
\noindent
Table~\ref{tbl:auto} presents the results.
To answer RQ2, we look at the orderings produced by each evaluation metric.  In terms of Reward, all simulation approaches produce the same ordering, which is the same as the one obtained with real users.  
The absolute values are also quite close to real users, with the exception of CIR6-PKG.  Since that approach terminates conversations earlier, it in a way models a more effective user behavior, which yields higher scores.
In terms of Success Rate, all but one simulator, CIR6-PKG, agree with the agent ranking produced by real users.  It should be noted that agents A and B are very close in terms of absolute scores, and CIR6-PKG in fact comes closest to real humans in terms of absolute numbers.  However, this method flips the order of agents A and B.  Let us point out that this is not unreasonable as A consistently ranked higher than B in terms of Reward.

It should be noted that these findings are based only on three systems.  Nevertheless, even with that, we can make some interesting observations that underline the need for further research on automatic evaluation measures.  
Looking at the relative orderings of agents across the two column in Table~\ref{tbl:auto}, it is clear that the two metrics disagree.  This is not a problem in itself, as they evaluate different aspects of conversational agents.  It, however, remains an open question which single metric aligns best with user satisfaction.

\shrink
\subsection{Realisticity}
\minishrink

\emph{(RQ3) Do more sophisticated simulation approaches (i.e., more advanced interaction and preference modeling) lead to more realistic simulation?}
Our working definition for a \emph{realistic} simulation is to be indistinguishable from conversations performed by real users. 

Specifically, we follow the multi-turn protocols defined in \citep{Li::2016:DRL} to compare a pair of dialogues conducted with a given conversational agent.  One of these dialogues is performed by a real user and the other is a simulated user.
Using crowdsourcing, we compare a sample of 25 simulated dialogues for each method and agent pair ($25 \times 3 \times 3 = 225$ dialogues in total).  
Each of the sampled dialogues is coupled with a human dialog with the corresponding agent and is shown side-by-side (in random order) to three workers on Amazon MTurk.  
Workers are then asked to choose which of the two dialogs was performed by a human. 
Ties are permitted when annotators find it difficult to distinguish.  
Additionally, workers are requested to give a brief explanation behind their choice. 
Options without explanations are filtered out.
We present the results in Table~\ref{tbl:human}.

To answer our research question, first we look at the effects of more advanced interaction modeling (QRFA-Single vs. CIR6-Single).
We find that our interaction model (CIR6) leads to substantially more wins (+6\% overall) over the existing model (QRFA).
Introducing personal knowledge graphs for preference modeling (CIR6-Single vs. CIR6-PKG) brings in further improvements (+3\% overall) in terms of wins.  We note that this is the best overall setting, even though not all agents benefit from this (specifically, agent \textbf{A}).

The results obtained using our best model (CIR6-PKG) are in fact quite remarkable, considering that 36\% of human evaluators have mistaken it for a real user, and 23\% of them could not decide whether it was a real user or not.

\minishrink
\section{Further Analysis}
\minishrink

\begin{table}[t]
	\caption{Classification of comments accompanying decisions when choosing which of two conversations was made by a real user versus a simulated one in a side-by-side comparison.  The columns are (Rea)listicity, (Eng)agement, (Emo)tion, (Res)ponse, (Gra)mmar, and (Len)gth.}
	\captionshrink	
	\begin{tabular}{lrrrlrrr}
	\toprule
	& \multicolumn{3}{c}{\textbf{Style}} && \multicolumn{3}{c}{\textbf{Content}} \\ 
	\cline{2-4} 	\cline{6-8}
	 &\textbf{Rea.} & \textbf{Eng.} & \textbf{Emo.} && 
	 \textbf{Res.} & \textbf{Gra.} & \textbf{Len.} \\
	\midrule
	QRFA-Single & 
		77 & 38 & 8 && 
		39 & 10 & 10  \\
	CIR6-Single & 
		76 & 31 & 8 && 
		53 & 15 & 3  \\
	CIR6-PKG & 
		74 & 33 & 15 && 
		34 & 14 & 9  \\
	\midrule
	All & 
		227 & 102 & 31 && 
		126 & 39 & 22 \\
	\bottomrule
	\end{tabular}
	\shrink
\label{tbl:anay}
\end{table}

Recall that in our last experiment crowd workers were tasked with deciding, in a side-by-side experiment, which of two dialogs were performed by a real user vs. a simulated one (a ``bot'').  In addition to making a simple choice, they were also asked to briefly explain their reasoning.  In this section, we further analyze these comments, in order to gain a better understanding of what traits of human behavior could be incorporated in a simulator in the future.

Based on an initial analysis of the comments, we came up with a coding scheme, which distinguishes between two main categories, dialogue Style and Content. 
We further subdivide Style into the following three categories:
(1) \textbf{Realisticity} is associated with how realistic or human-so\-unding a dialog is. For example, ``\emph{User 2 seems a bit more human-like and realistic}'' and ``\emph{The user is more genuine and stubborn about his requests which seem very natural}.''
(2) \textbf{Engagement} is about the involvement of the user in the conversation, e.g., ``\emph{There appears to be more attempts at dialogue in the second conversation that seems more human}'' and ``\emph{The first one is authentic and adds their opinion to each statement}.''
(3) \textbf{Emotion} refers to expressions of feelings or emotions. For example, ``\emph{The user in dialogue 1 shows emotions like when he shows how he loves mila}'' and ``\emph{Dialogue 2 expresses feeling, robots have no feelings}.''

We also distinguish between three Content subcategories:
(1) \textbf{Response} refers to cases where the user does not seem to understand the agent correctly (``\emph{some badly answered answers}'') or repetitively asks the same question (``\emph{the first one was too repetitive}'').
(2) \textbf{Grammar} is about language usage, including spelling and  punctuation. For example, ``\emph{User 1 made a spelling mistake with anna karinina, which i doubt a bot would do},'' and ``\emph{The user makes a typo in the left dialog}.''
(3) \textbf{Length} concerns the length of reply (``\emph{Very short and simple to the point}'') or of a conversation (``\emph{they had a longer conversation}'').

Table~\ref{tbl:anay} presents the statistics.  Note that the numbers here do not mean success or failure; they merely indicate how often each aspect was considered when deciding whether the user in the conversation was a human or a bot.  We hypothesize that the biggest gains in creating more human-like simulations lie in improving the aspects that were mentioned most.
These are, in order: Realisticity (41\%), Response (23\%), and Engagement (19\%), where the percentages are calculated with respect to all annotations.  
To improve Realisticity, one could imagine using a more natural tone for expressing preferences.
The Response aspect may be enhanced by keeping a better track of conversation history and by generating more varied responses.
As for Engagement, we observed that humans tend to continue the discussion and explore the space of options, even after they have found a recommendation they liked. 
\section{Conclusions and Future Directions}

We have introduced a simulation framework that enables large scale automatic evaluation of conversational recommender systems. 
Our simulation approaches, equipped with a preference model, interaction model, NLG, and NLU, are capable of generating human-like responses. 
We evaluate them by comparing three existing conversational movie recommender systems.
The results indicate that the preference model and task-specific interaction models can achieve high correlation between automatic and human evaluations.

This work represents a first important step towards evaluating conversational information access systems using simulation. 
We see a number of directions for extending it in future work. 
First, we wish to generalize our findings by conducting simulations in multiple domains.  Even though nothing is domain specific in our approach, the limited availability of conversational services in other domains represents a challenge. 
Second, there is a lack of automatic evaluation metrics for conversational information access.  Our results indicate that it is possible to obtain the same relative ranking of agents via simulation than with real users for a given metric. However, current evaluation metrics do not agree with each other on how to rank agents.  Thus, an important future research objective is to develop evaluation metrics that better align with user expectations and satisfaction.  
Additionally, we see potential in using simulation to debug conversational agents.  For example, one could identify the type of actions where the system fails or locate points in a sequence of dialog turns when a given performance metric tends to drop.  Then, by providing a sequence of utterances up to a given point, possible continuations of the dialog could be evaluated from that point.
Third, some of our components rely on hand-crafted heuristics or make simplifying assumptions.  
For example, our NLG generates natural language responses with a template-based model.
In the future, we wish to improve each individual component by investigating advanced natural language processing techniques.  For example, for NLG we could using deep learning methods to learn to generate more human-like responses.
Finally, motivated by the insights from our side-by-side evaluation, we wish to equip our simulated users with more personalized traits, such as emotion, engagement, and patience.

\bibliographystyle{ACM-Reference-Format}
\bibliography{00paper}

\end{document}